\documentclass[%
 reprint,
 amsmath,amssymb,
 aps,
]{revtex4-2}

\usepackage{graphicx}% Include figure files
\usepackage{dcolumn}% Align table columns on decimal point
\usepackage{bm}% bold math
\begin{document}

\preprint{APS/123-QED}

\title{Beating Poisson stochastic particle encapsulation in flow-focusing microfluidic devices using viscoelastic liquids}% Force line breaks with \\

\author{Keshvad Shahrivar}
\author{Francesco Del Giudice}%
 \email{francesco.delgiudice@swansea.ac.uk}
\affiliation{%
 Department of Chemical Engineering, School of Engineering and Applied Science, Faculty of Science and Engineering, Swansea University, Swansea, SA1 8EN
}%


%\affiliation{
%\affiliation{
 %Third institution, the second for Charlie Author
%\affiliation{%


\date{\today}% It is always \today, today,

\begin{abstract}
The encapsulation and co-encapsulation of particles in microfluidic flows is essential in applications related to single-cell analysis and material synthesis. However, the whole encapsulation process is stochastic in nature, and its efficiency is limited by the so-called Poisson limit. We here demonstrate particle encapsulation and co-encapsulation in microfluidic devices having flow-focusing geometries with efficiency up to 2-folds larger than the stochastic limit imposed by the Poisson statistics. To this aim, we exploited the recently observed  phenomenon of particle train formation in viscoelastic liquids, so that particles could approach the encapsulation area with a constant frequency that was subsequently synchronised to the constant frequency of droplet formation. We also developed a simplified expression based on the experimental results that can guide optimal design of the microfluidic encapsulation system. Finally, we report the first experimental evidence of viscoelastic co-encapsulation of particles coming from different streams.

%\begin{description}
%use the optional argument of the \verb+\item+ command to give the %category of each item. 
%\end{description}
\end{abstract}

\maketitle

%\tableofcontents
The compartmentalisation of particles in picoliter droplets obtained using microfluidic devices finds application across several fields of science~\cite{del2021microfluidic}, with some examples being single-cell analysis~\cite{macosko2015highly} and material synthesis~\cite{turek2016self,yi2003generation}. The compartmentalisation process is driven by the same principle of droplets formation in microfluidic devices: a continuous phase `meet' at a junction the dispersed phase containing particles or cells, and a droplet is formed as a result of a combination among interfacial, viscous, inertial and elastic forces (to form the droplet, the two fluids must not be miscible with each other)~\cite{nunes2013dripping}. The microfluidic junction can present different shapes, for instance coaxial, T-shaped or flow-focusing~\cite{garstecki2006formation,nunes2013dripping}. The process of encapsulating objects in droplets is stochastic and it is governed by the Poisson limit~\cite{kemna2012high}. The stochastic nature of the encapsulation process finds its origin in the fact that objects approaching the encapsulation area do not arrive with a constant frequency, at variance with the droplet formation process where the frequency of droplet formation is constant and can be tuned by changing the values of the volumetric flow rate for both continuous and dispersed phase~\cite{nunes2013dripping,du2016breakup}. To `beat' the stochastic encapsulation limit (that for single particle encapsulation is at around 30\%, while for co-encapsulation is around 10\%), it is possible to take advantage of the formation of strings of equally-spaced particles (also called particle trains) in microfluidic geometries thanks to either inertial~\cite{matas2004trains,humphry2010axial,kahkeshani2016preferred,lee2010dynamic,gao2017self,dietsche2019dynamic,hu2020stability,liu2022} or viscoelastic~\cite{d2013dynamics,del2018fluid,davino2019pairs,d2020numerical,jeyasountharan2021viscoelastic} forces. In this way, the constant frequency of particles approaching the encapsulation area $f_p$ can be synchronised to the frequency of droplet formation $f_d$, i.e., $f_p=f_d$, to overcome the Poisson stochastic limit. This approach has been successfully used in inertial microfluidics~\cite{edd2008controlled,li2019dean,harrington2021dual} for a variety of microfluidic channel designs. Very recently, we employed a viscoelastic shear-thinning fluid to demonstrate the controlled encapsulation of rigid particles in a T-junction microfluidic device taking advantage of the viscoelasticity-driven particle train formation phenomenon~\cite{shahrivar2021controlled}. So far, this have been the only study on the subject and it was restricted to single particle encapsulation in a commercially available microfluidic device.

In this letter, we demonstrate controlled viscoelastic encapsulation and co-encapsulation of particles in flow-focusing microfluidic devices, achieving encapsulation and co-encapsulation efficiency up to 2-folds larger than the stochastic value, thus beating the Poisson limit. In practical terms, this work represents an extension of our previous one~\cite{shahrivar2021controlled} featuring controlled viscoelastic encapsulation in T-junction devices.

\begin{figure*}[t!]
\includegraphics[scale=1]{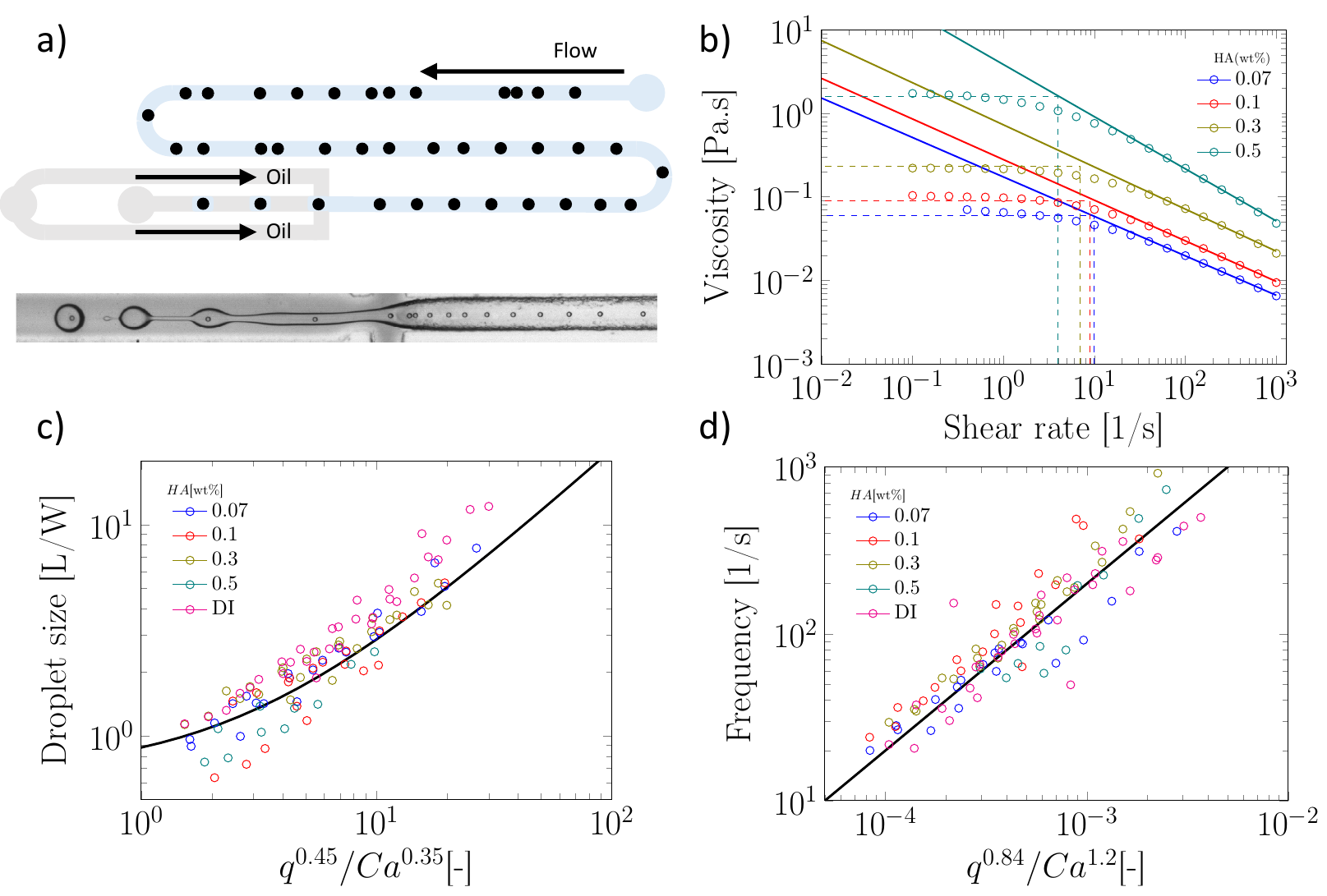}% Here is how to import EPS art
\caption{\label{fig:SchematicDroplet} a) Schematic of the microfluidic encapsulation device together with an experimental snapshot of the encapsulation process. b) Viscosity curve for several HA aqueous solution obtained using a conventional rheometer (AR2000ex, TA instruments) equipped with a 60~mm 1$^\circ$ cone and plate geometry. c) Scaling of droplet size $L$ normalised by the lateral channel width $W$ as a function of the ratio the flow rate ratio $q$ and the Capillary number $Ca$. d) Scaling of frequency of droplet formation as a function of the flow rate ratio $q$ and the Capillary number $Ca$.}
\end{figure*}

We first demonstrated the principle of controlled viscoelastic encapsulation using the flow-focusing microfluidic device schematised in Fig.~\ref{fig:SchematicDroplet}a. 
The suspension containing particles flowed in a polymethylmethacrylate (PMMA) square-shaped microchannel with lateral side $H=100~\mu$m, width $W=100~\mu$m and length of 30~cm, bonded on a glass coverslip using double-sided tape (Adhesive Research). The channel length was selected such that the $20\pm 2~\mu$m particles (Polysciences inc) employed here at a bulk concentration of 0.4~wt\% self-ordered in a particle train thanks to the viscoelastic properties of the suspending liquid~\cite{del2018fluid,jeyasountharan2021viscoelastic} before approaching the encapsulation area. The width of the section following the encapsulation area was $W=120~\mu$m, while the height remained unchanged at $H=100~\mu$m. Mineral oil (Sigma Aldrich UK, viscosity of 29~mPa$\cdot$s) was used as the continuous phase to form viscoelastic droplets. The PMMA microchannel was fabricated using the same procedure reported elsewhere~\cite{del2018fluid}. As viscoelastic suspending liquid, we selected several aqueous polymer solutions of hyaluronic acid (HA, molecular weight $M_w=1.5-1.8$~MDa, Sigma Aldrich, UK) displaying different amount of shear-thinning (Fig.~\ref{fig:SchematicDroplet}b). The fluids displayed strong elastic properties (Fig.S1), with longest relaxation time values estimated from the viscosity data of Fig.~\ref{fig:SchematicDroplet}b, by determining the intersection between the power-law region and the zero-shear plateau region of the curve. Relaxation time values of 0.10, 0.11, 0.143 and 0.25~s were obtained for the HA concentrations 0.07, 0.1, 0.3 and 0.5wt$\%$, respectively. The interfacial tension values between the HA solutions and the mineral oil measured using a force tensiometer (Sigma 702, biolin scientific) were $\gamma=3.63\pm 0.07, 3.64\pm 0.05, 3.04 \pm 0.16$, and $1.65 \pm 0.08$~mN/m for the HA concentrations 0.07, 0.1, 0.3 and 0.5wt$\%$, respectively. We used a fast camera (Photron Mini UX50) mounted on a inverted Zeiss Axiovert A1 microscope; videos of recorded droplets were analysed using a Matlab subroutine employed in our previous study~\cite{shahrivar2021controlled}. The flow rate of the two liquids were controlled using a syringe pump (KD Scientific). 

\begin{figure}[t!]
\includegraphics[scale=0.65]{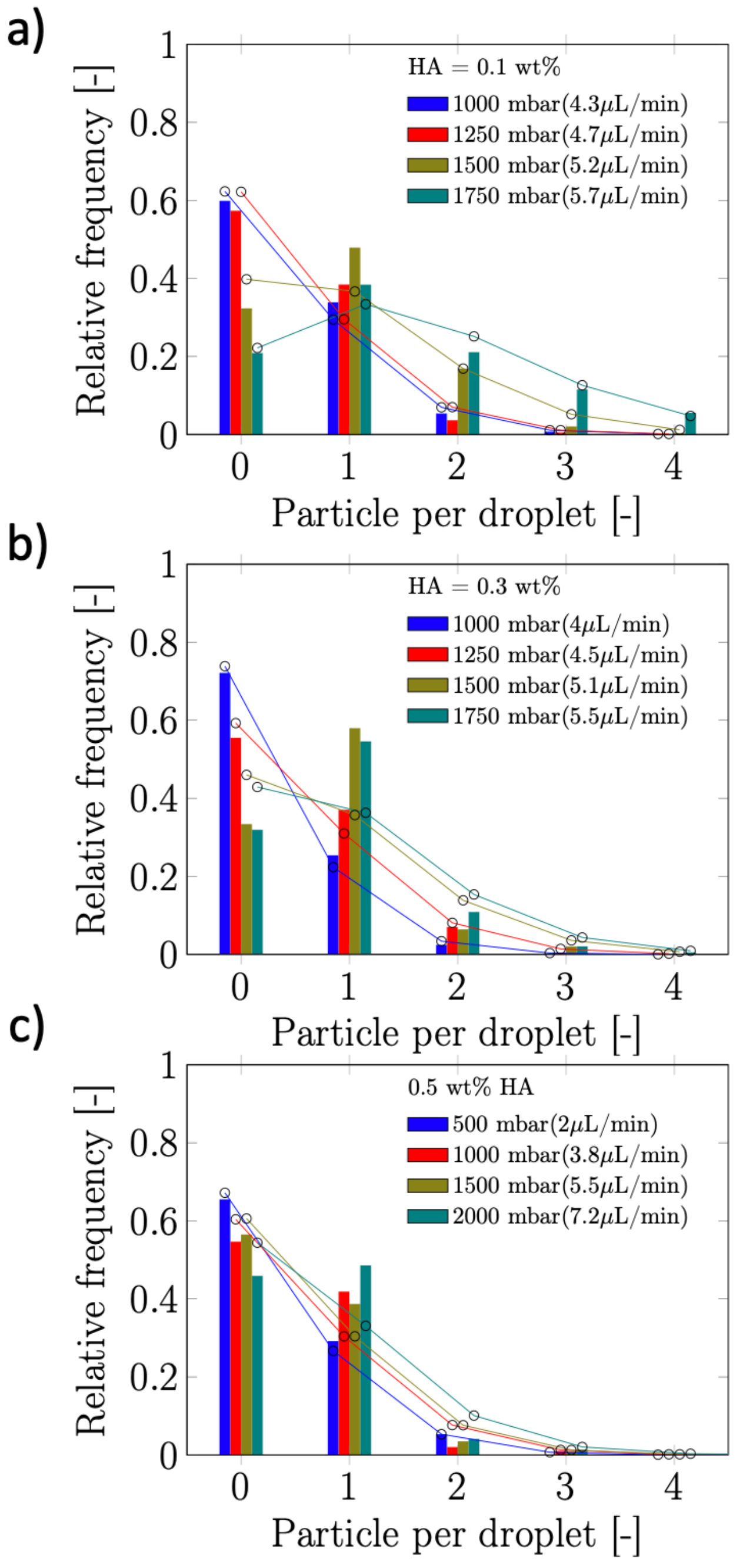}
\caption{\label{fig:Encap} Histograms of relative frequency of encapsulated particles as function of the particles per droplets at different imposed pressure drops of the dispersed phase (controlled using a Dolomite p-pump) and flow rate of the continuous phase $Q_c=7~\mu$L/min, for a) HA~0.1~wt\%, b) HA~0.3~wt\%, c) HA~0.5~wt\%. Open symbols connected by solid lines represent the Poisson statistic limit experimentally predicted for stochastic encapsulation.}
\end{figure}

Similarly to our previous work~\cite{shahrivar2021controlled}, we first studied the droplet formation phenomenon (Fig.~\ref{fig:SchematicDroplet}(c-d)) to verify whether the experimental data could be described using a single scaling argument. The droplet formation phenomenon generally depends upon a series of parameters~\cite{garstecki2006formation,nunes2013dripping}, including the flow rate ratio $q=Q_d/Q_c$, where $Q_d$ is the flow rate of the dispersed phase (HA solutions) and $Q_c$ that of the continuous phase (Mineral oil), and the Capillary number $Ca=\mu_c U_c/\gamma$, where $U_c=Q_c/(W H)$ is the average velocity of continuous phase fluid, $\mu_c$ is the viscosity of the continuous phase and $\gamma$ is the interfacial tension. In our experiments, the flow rate $Q_c$ was changed in range of $0.1-10~\mu$L/min corresponding to capillary number in the range of $Ca \in$ [0.001 - 0.1]. For this range of control parameters, the largest value of the Reynolds number of the continuous phase $Re = \rho_c U_c W/\mu_c$ was $Re=6.4\times10^{-2}$, hence, inertial effects were negligible. We found that, regardless of the polymer concentration, the viscoelastic droplets collapsed on the mastercurve given by the expression recently reported by Chen~et~al~\cite{chen2020modeling}, $L/W=a+\epsilon(q^b/Ca^c)$, with fitting parameters $a=0.66\pm 0.02$, $b=0.45\pm0.01$ and $c=0.35\pm0.01$, respectively (Fig.~\ref{fig:SchematicDroplet}c). It is worth noticing that the expression by Chen~et~al.~\cite{chen2020modeling} was developed for Newtonian droplets surrounded by a continuous shear-thinning phase, a somehow opposite situation to the one investigated here. In other words, the fact that the data followed the scaling for Newtonian droplets may indicate that the droplet size is not significantly affected by the fluid elasticity. We observed a similar phenomenon in our recent work featuring viscoelastic droplets in T-junction microfluidic devices~\cite{shahrivar2021controlled}, thus suggesting that fluid viscoelasticity impacts more the dynamics of droplet formation rather than the final size, in agreement with other studies~\cite{rostami2018generation,fatehifar2021non,wong2017numerical,ren2015breakup,du2016breakup,chen2021pressure,xue2020breakup}. Similarly to the droplet size, we also observed that the frequency of droplet formation $f_d$ scaled according to a mastercurve dependent upon the flow rate ratio $q$ and the Capillary number $Ca$ (Fig.~\ref{fig:Encap}d), with expression: %We observed the same phenomenon in our previous work~\cite{shahrivar2021controlled}. Here the functional form of the curve could be made dimensionless according to the following expression:
\begin{equation}
f_d=\alpha\frac{q^{m}}{Ca^{n}},
\label{eq_fd}
\end{equation}
where $\alpha=2.00\pm0.02\times 10^{5}$~1/s, $m=0.84\pm 0.05$ and $n=1.20\pm 0.05$, obtained via a standard fitting. %The scaling is also in good agreement with recent experiments in flow-focusing geometries where the frequency of droplet formation displayed a nearly linear variation with the Capillary number~\cite{chen2021pressure}.

The results presented lay the foundation to demonstrate the controlled encapsulation principle in flow-focusing geometries thanks to the viscoelasticity of the dispersed suspending phase. When the process of loading particles into drops is purely random, the Poisson statistics predicts that the probability $P$ of droplets containing $n$ particles is given by:
\begin{equation}
P=\frac{k^{n}\exp(-{k})}{n!}, 
\label{eq_ProbEnap}
\end{equation}
where $k$ is the average number of particles per droplet. For the encapsulation to be controlled, i.e., with encapsulation efficiency above the Poisson stochastic limit, it is essential that particles approaching the encapsulation area arrive with a constant frequency, a condition that is only possible when particles self-assemble in a train of equally-spaced particles~\cite{shahrivar2021controlled}. We here exploited both the viscoelastic and shear-thinning properties of the HA solutions to control the spacing between consecutive particles in order to demonstrate controlled single particle encapsulation. We observed that particles arrived at the encapsulation area simultaneously focused at the centreline and with a preferential spacing so that could be subsequently encapsulated in a controlled manner (experimental snapshot in Fig.~\ref{fig:SchematicDroplet}a).
Owing the combination of the viscoelasticity-driven particle ordering and droplet formation, we observed a single particle encapsulation efficiency value above the Poisson stochastic limit for all the viscoelastic fluids employed (Fig.~\ref{fig:Encap}(a-c)). For each HA concentration, we observed a non-monotonic behaviour for the encapsulation efficiency, with a unique combination of parameters in which the single encapsulation efficiency was the highest possible. This is clearly expected as the optimal encapsulation conditions are only observed when $f_d=f_p$. Please note that the frequency of droplet formation, $f_d$, changes with different values of the flow rate of the dispersed phase~\cite{garstecki2006formation,nunes2013dripping}, while the frequency of particle approaching the encapsulation area, $f_p$, does not depend significantly on the volumetric flow rate~\cite{del2018fluid,jeyasountharan2021viscoelastic}. This is also in good agreement with our previous work on viscoelastic particle encapsulation in T-junction~\cite{shahrivar2021controlled} and other works on inertial encapsulation~\cite{edd2008controlled}. We also derived an expression that can guide the choice of experimental parameters aimed at obtaining a controlled encapsulation. In an ideal particle train with local concentration $\phi_l=Nd/L$, where $N$ is the number of particle observed in a video frame, $d$ is the particle diameter and $L$ is the length of the video frame, particles are equally spaced with a constant inter-particle distance~\cite{del2018fluid,jeyasountharan2021viscoelastic}. The expression for the frequency of particles approaching the encapsulation area is $f_p=u \phi_l/d$~\cite{shahrivar2021controlled}, where $u$ is the particle velocity. The use of $\phi_l$ over the bulk particle concentration is motivated by the fact that $\phi_l$ accounts for fluctuations in the local particle concentration, while the bulk particle concentration does not~\cite{jeyasountharan2021viscoelastic,shahrivar2021controlled,kahkeshani2016preferred}. (The relation between $\phi_l$ and the bulk concentration $\phi_b$ is~\cite{del2018fluid,shahrivar2021controlled} $\phi_b=\frac{2}{3}\phi_l \beta^2$, where $\beta=d/H$ is the confinement ratio for square-shaped geometries). By combining the expression for $f_p$ with the expression for $f_d$ (Eq.~\ref{eq_fd}), the optimal encapsulation condition ($f_d=f_p$) is:
\begin{equation}
%Q_c = Q_d Ca^{\frac{n}{m}}\left(\frac{\alpha \pi d \beta^{2}}{6 u \phi} \right)^{\frac{1}{m}}\\
q=\left(\frac{u \phi_l}{\alpha d}\right)^\frac{1}{m} Ca^{\frac{n}{m}}.
\label{eq_OptEncap}
\end{equation}
Being $n\approx m\approx 1$ (from the fitting of Eq.~\ref{eq_fd}), an approximated expression of Eq.~\ref{eq_OptEncap} is:
\begin{equation}
    q=\frac{u \phi_l}{\alpha d} Ca.
    \label{eq_OptEncapApprox}
\end{equation}
Note that Eq.~\ref{eq_OptEncapApprox} is very similar to the equation we previously found in T-junction geometry, when using xanthan gum as viscoelastic liquid~\cite{shahrivar2021controlled}. 

\begin{figure}[t!]
\includegraphics[scale=1]{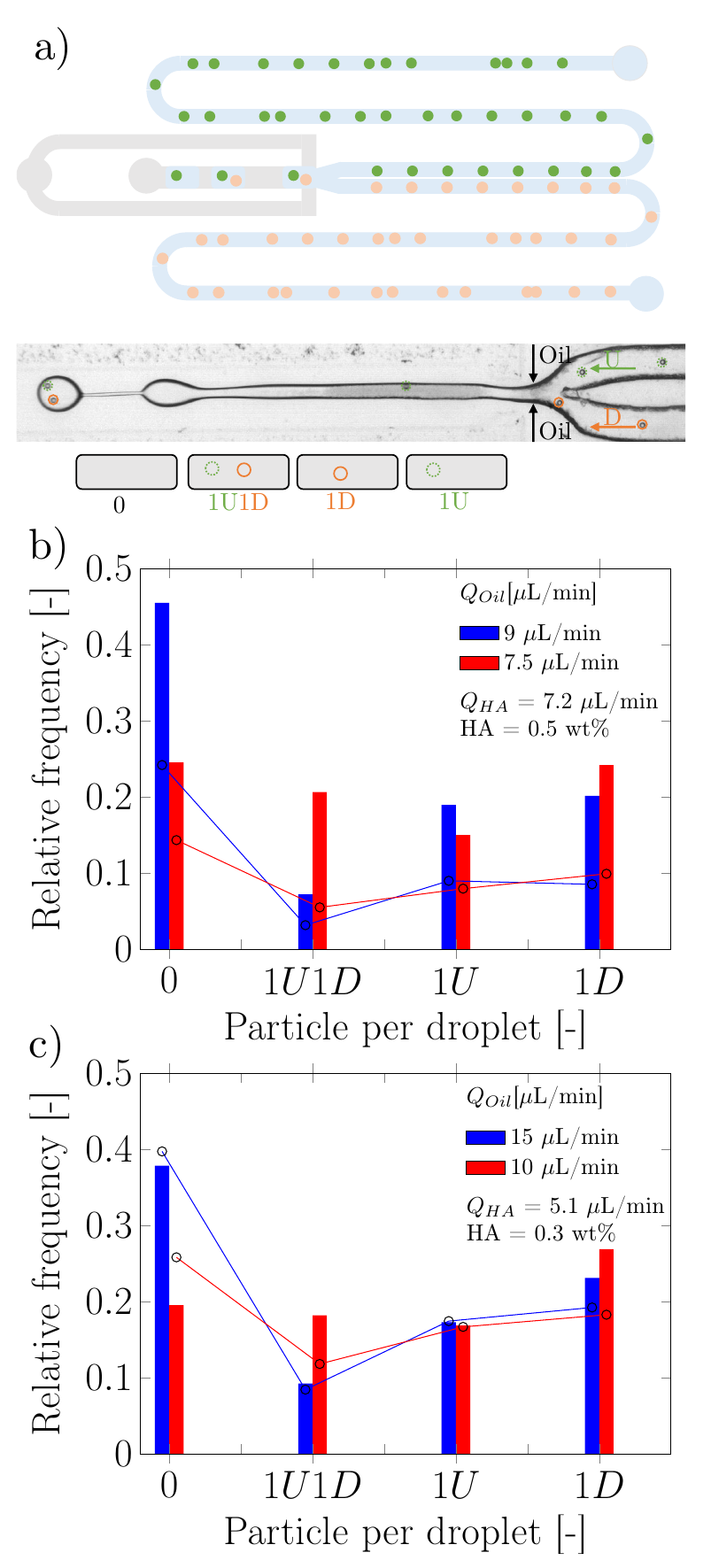}
\caption{Co-encapsulation of particles in a flow-focusing device. a) A schematic representation of the microchip. Particles type $U$ from up channels meet particles type $D$ from down channel at the junction and co-encapsulate into HA droplets. b) Probability of co-encapsulation for HA 0.5 wt$\%$. c) Probability of co-encapsulation for HA 0.3 wt$\%$. At least 650 droplets were analysed to obtain statistics. Open symbols connected by solid lines represent the Poisson statistic limit experimentally predicted for stochastic co-encapsulation.}
\label{fig:CoEncap}
\end{figure}

While single particle encapsulation is important in the context of optimising compartmentalisation efficiency~\cite{del2021microfluidic}, a step further aimed at targeting single-cell analysis applications is the co-encapsulation of particles and cells arriving from two different flow streams. For instance, this is extremely relevant for Drop-seq applications~\cite{macosko2015highly}, where flowing cells need to be encapsulated together with functionalised beads in the same droplet in order to extract the genome of the cell. This application certainly benefits from a controlled encapsulation perspective, as the number of unused beads decreases while the number of analysed cells in a single run increases~\cite{li2019dean}. Motivated by the success in achieving a controlled single particle encapsulation in a flow-focusing device, we investigated the possibility of overcoming the Poisson stochastic limit in co-encapsulation applications using viscoelastic liquids. Here, the co-encapsulation of particles was studied by introducing a second inlet to the original single encapsulation design (see Fig.~\ref{fig:CoEncap}a).
The two stream of dispersed phase, here referred as type $U$ and type $D$ (representing "up" and "down" stream from a top view of the device), met at the junction, while the continuous mineral oil phase did cut the HA dispersed phase to form viscoelastic droplets containing particles coming from the two streams. The Poisson probability for the co-encapsulation in a droplet containing $n_D$ particles of type $D$ and $n_U$ particles of type $U$ is given by the product of the two independent Poisson probabilities as: 
\begin{equation}
P =  \frac{k_U^{n_U}\exp(-{k_U})}{n_U!} \times \frac{k_D^{n_D}\exp(-k_{D})}{n_D!},
\label{eq_ProbCoEnap}
\end{equation}
here $k_{U}$ and $k_{D}$ are the average number of particles per droplet for particles  type $U$ and type $D$, respectively. The Poisson stochastic value obtained when the average number of particles per droplet is equal to 1 (i.e., ${k_U}= {k_D}=1$) is 13.5$\%$ for one-to-one co-encapsulation (i.e., ${n_U}={n_D}=1$). We demonstrated that viscoelastic fluids were successful in increasing the co-encapsulation efficiency up to 2-folds the Poisson limit (Fig.~\ref{fig:CoEncap}(b-c)). For HA 0.5wt$\%$ and at constant dispersed phase flow rate $Q_d=7.2\mu$L/min, we observed that a continuous flow rate of the dispersed phase $Q_c=7.5\mu$L/min resulted in a one-to-one co-encapsulation efficiency ($1U1D$) of 22\%, well above the Poisson limit (open symbol on the bars in Fig.~\ref{fig:CoEncap}b). A similar trend was observed for HA 0.3wt$\%$, with a  one-to-one co-encapsulation efficiency of 20\% for $Q_c=10\mu$L/min and $Q_d=5.2\mu$L/min. These efficiency values were obtained by counting at least 650 droplets. This is the first experimental evidence of viscoelastic co-encapsulation of particles in microfluidic devices, being the existing literature on the topic of controlled co-encapsulation focusing on inertial flow~\cite{lagus2013high,li2019dean}.

In summary, we here demonstrated controlled encapsulation and co-encapsulation of particles in flow focusing geometries using viscoelastic fluids, achieving efficiency up to 2-folds larger compared to the Poisson value. The results presented here are an extension of our previous work on T-junction geometries~\cite{shahrivar2021controlled}. Future works should extend our findings further to include different particle diameters as well as experiments on cells. With regards to the second point, this is not straightforward as cells require phosphate buffer saline (or cell media) to survive in solution: being HA a polyelectrolyte, addition of salt to a fixed polymer concentration would alter the rheological properties by decreasing both the elasticity and the shear-thinning properties~\cite{del2017relaxation,colby2010structure}: both conditions are required to observe self-assemble of viscoelastic particle trains in microchannels~\cite{del2018fluid}. 

The results presented in this manuscript have also been submitted as part of a patent application. F.D.G acknowledges support from EPSRC New Investigator Award, grant EP/S036490/1. The data that support the findings of this study are available from the corresponding author upon reasonable request.

%\begin{acknowledgments}

%\end{acknowledgments}

\nocite{*}

%\bibliography{BeatingPoissonBIB}% Produces the bibliography via BibTeX.

\providecommand{\noopsort}[1]{}\providecommand{\singleletter}[1]{#1}%

\end{document}